# Post-melting encapsulation of glass microwires for multipath light waveguiding within phosphate glasses

I. Konidakis*[1], F. Dragosli[1], A. Cheruvathoor Poulose[2], J. Kašlík[2], A. Bakandristos[2,3], R. Zboril[2,3] and E. Stratakis[1]

1. Institute of Electronic Structure and Laser (IESL), Foundation for Research and Technology-Hellas (FORTH), 70013 Heraklion-Crete, Greece.
2. Regional Centre of Advanced Technologies and Materials, Czech Advanced Technology and Research Institute (CATRIN), Palacký University, Šlechtitelů 27, 783 71, Olomouc, Czech Republic.
3. Nanotechnology Centre, Centre of Energy and Environmental Technologies, VŠB-Technical University of Ostrava, Ostrava-Poruba, Czech Republic.

**Abstract**

Glass waveguides remain the fundamental component of advanced photonic circuits and with a significant role in other applications such as quantum information processing, light generation, imaging, data storage, and sensing platforms. Up to date, the fabrication of glass waveguides relies mainly on demanding chemical processes or on the employment of expensive ultrafast laser equipment. In this work, we demonstrate the feasibility of a simple, low-temperature, post-melting encapsulation procedure for the development of advanced glass waveguides. Namely, silver iodide phosphate glass microwires (MWs) are drawn from typical splat-quenched samples. Following this, the MWs are incorporated in a controlled manner within previously prepared transparent silver phosphate glass rectangular prisms. The composition of the employed glasses is chosen so that the host phosphate glass has a lower refractive index than the embedded MWs. In such case, the waveguide mechanism relies on the propagation of light inside the encapsulated higher refractive index MWs. Moreover, the presence of silver nanoparticles within the MWs enhances the light transmission due to scattering effects. Waveguide devices with either one or two incorporated MWs were fabricated. Remarkably, in the latter case, the transmission of light of different colors and in multipath direction is possible, rendering the developed waveguides outstanding candidates for various photonic circuits, optoelectronic, and smart sign glass applications.

*Corresponding author: ikonid@iesl.forth.gr

**Keywords:** phosphate glass, multipath waveguides, glass microwires, silver nanoparticles, post-glass melting encapsulation.

# 1. Introduction

Controlling light guidance throughout transparent solids remains a continuous scientific and technological challenge in the fields of photonics and optoelectronics. In recent years many optical materials were employed in the fabrication of integrated photonic circuits and advanced waveguides, and glass remains a fundamental one among them [1-3]. Apart from the extensive use in photonic circuits, glass waveguides expand to other numerous applications that include photodetection [4], light generation photonic devices [5], random laser and imaging [6], data storage components [7,8], quantum information processing [9], and sensing platforms [10,11].

The fabrication of glass optical waveguides relies on two major approaches, a first that involves thin film deposition procedures, and a second that is based on the glass refractive index modification [12]. The thin film deposition itself is accomplished by physical or chemical processes. Physical processes involve demanding vacuum techniques such as electron beam evaporation [12], and pulsed laser deposition [13,14], as well as sputtering methods such as magneton [15], and ion beam sputtering [16]. Chemically driven thin film deposition methods could be even more demanding, as they require gas phase deposition, like vapor [17,18], and plasma deposition approaches [12], or liquid phase protocols, like sol-gel [19,20], and spray pyrolysis processes [12]. Rather differently, the glass index modification is achieved by means of ion exchange [21], as well as, by UV and ultrafast laser patterning of the glass [2,3,22.23]. Overall, and despite the efforts made over the years, it becomes apparent that the fabrication of glass waveguides still relies on the employment of expensive laser and optical equipment, or on demanding, and in some cases environmental unfriendly techniques for the implementation of coatings and thin films.

In recent studies we have shown the excellent feasibility of a post-melting encapsulation procedure towards the development of ultrastable and highly luminescent perovskite glasses [24], composite two-dimensional (2D) materials glass nanoheterojunctions [25], and superior photochromic glasses [26]. Herein we exploit the post-melting protocol for the development of advanced all-glass waveguides without the need for laser processing or chemical procedures. In particular silver iodide phosphate glass microwires (MWs) are drawn from typical splat-quenched samples. Following the drawing, the glass MWs are incorporated in a controlled manner within previously prepared transparent silver phosphate glass rectangular prisms. Notably, the MWs glass exhibits a higher refractive index than the host glass, allowing the controlled light propagation within the waveguide devices by means of total internal reflection (TIR) principles. Moreover, the proposed simple, low-temperature, post-melting procedure allows the incorporation of more than one MWs within the host glass, rendering possible the light transmission of different colors and in multiple directions. Based on this, the fabricated devices are suitable candidates for advanced photonic circuits, optoelectronic platforms, and smart sign applications.

## 2. Experimental

### 2.1 Development of glasses, glass microwires (MWs), and waveguide devices

For the development of the waveguides, two silver phosphate glasses within the $xAgI+(1-x)AgPO_3$ family were synthesized. Namely, the binary silver phosphate glass ($AgPO_3$) with x=0 was employed as the host waveguide glass, whereas the silver iodide phosphate glass with x=0.3 was used for the MWs drawing. Both glasses were prepared by melting appropriate amounts of AgI, $AgNO_3$, and $NH_4H_2PO_4$ dry powders within an electrical furnace upon following a typical procedure described explicitly in our previous works [24-27]. $AgPO_3$ glasses were obtained in the form of rectangular prism blocks of various top area dimensions (0.5 x 1 cm$^2$ and 1 x 1 cm$^2$), after casting the melt inside custom made moulds. For improving the optical quality of the upper surface, the glasses were stamped with an optically polished silicon wafer. Fig. S1 depicts indicative samples. The x=0.3 glasses were splat-quenched between two silicon wafers in the form of 1 mm thick discs with a diameter of 1 cm$^2$. The inset of Fig. 1 presents a photograph of a typical glass disc to be used for the development of MWs.

For the MWs drawing, an as-prepared splat-quenched $0.3AgI+0.7AgPO_3$ glass was placed on a silica wafer positioned on a heating plate. The glass was heated up to 150 ºC. As the glass approaches the glass transition temperature ($T_g$) of 148 ºC [27], it gains viscosity. Immediately after, a tip of a typical glass pipette was emersed within the sample and moved upwards as depicted schematically in Fig. 1a. The melted glass sticks to the tip of the pipette, and as the pipette moves away from the splat-quenched glass a MW was drawn as demonstrated in Vid. S1. The MW is then removed from the end face of the pipette, whereas the speed of removing the pipette from the glass determines the MW thickness.

After drawing, the $0.3AgI+0.7AgPO_3$ glass MWs were cleaved to the appropriate length in order to be encapsulated within the host binary $AgPO_3$ glass. As shown in Fig. 1b the host glass is positioned on a silica wafer placed on a heating plate. The inset of Fig 1b depicts photos of typical $AgPO_3$ glass rectangular prisms that are employed for the fabrication of the waveguide devices. On the upper surface of the host glass, the MW is placed at the intended immersing position as shown in Fig. 1b. The temperature rises up to 190 ºC, i.e. close to the $T_g$ of the binary glass (192 ºC) [24]. Once the host glass gains viscosity the MWs is pressed from the top so that is totally immersed within the host glass (Fig. 1b). Immediately after, the silicon wafer is removed from the heating plate, and the waveguide composite glass device cools down to room temperature. Notably, the same route applies for single and multipath waveguide devices were more than one MWs are incorporated at different directions within the same host phosphate glass prism.

**2.2 Materials characterization**

A scanning electron microscope (JEOL, JSM-7000F) was employed for the morphological examination of the synthesized glasses and the waveguide devices, along with scanning transmission electron microscopy (STEM) studies in the high-angle annular dark field (HAADF) mode. The elemental mapping was performed with an FEI TITAN G2 60-300 HRTEM microscope with an X-FEG type emission gun, operating at 300 kV, objective-lens image spherical aberration corrector and ChemiSTEM EDS detector. For the optical properties, ultraviolet-visible (UV–Vis) absorption spectra were collected on a Cary 50 UV-Vis spectrophotometer (Varian). The structural modifications of the synthesized glasses were studied by means of Raman spectroscopy. Namely, room temperature Raman spectra with a resolution of 1 $cm^{-1}$ were collected at the backscattering geometry upon employing a 532 nm laser line for excitation [24,25].

The waveguiding features of the developed single MW glass waveguides were demonstrated upon employing three distinct light emission sources. In particular, a blue cw laser emitting at 450 nm, a green cw laser emitting at 526 nm, and a red cw laser emitting at 680 nm were used. Fig. S2a presents the experimental set up with the necessary optical components to guide the selected beam towards the waveguide device at one at a time manner. A microscope objective mounted on an x, y, z stage was employed to assist the coupling of light within the encapsulated MW of the waveguide (Fig. S2a). For the multipath waveguide devices, the green and red laser were used. The green laser is coupled within the parallel to the edges of the glass MW as shown for the single MW devices in Fig. S2a. In addition, a convex lens from the opposite direction is employed to focus the red laser beam on the second MW (Fig. S2b), which is incorporated parallel or diagonally to the other MW that waveguides the green light throughout the glass. A typical power meter was used to determine the transmission losses of the fabricated waveguides, upon comparing the power of the incident light with that of the output light on the other side of the waveguides (Fig. S2a).

**3. Results and discussion**

Fig.2a presents a scanning electron microscopy (SEM) image of the splat-quenched $0.3AgI+0.7AgPO_3$ glass that was used for drawing the MWs. Figs. 2b-d depict typical SEM top-view photos of a drawn MW with a diameter of 100 µm. It becomes apparent that the employed method results to the formation of uniform glass MWs with smooth and homogeneous surface. The same applies to thinner curved MWs (Fig. 2e), that were drawn upon following the same procedure. Fig. 2f shows the upper surface of the $AgPO_3$ host glass upon encapsulation of the glass MW, whereas Fig. 2g presents a magnified area. Inspection of Fig. 2g reveals that the MW is immersed within the host glass at a depth of around 200 µm, while leaving a notable mark on the surface of the host glass block. A typical SEM micrograph of the immersion point across the surface of the host glass is depicted in Fig. 2h. Notably, following the incorporation of the glass MW, the surface of the host glass remains smooth. As it is depicted in Fig. 2f, at the upper side of the waveguide device the incorporated glass MW has been cleaved exactly on the end face of the glass to facilitate light coupling. Rather differently, at the output side of the device, as pointed out by the red arrow in Fig. 2f, the encapsulated glass MW was cleaved at an exceeding length of around 80 µm, in order to demonstrate better and monitor the light transmission through it, when the waveguiding features of the devices are studied.

It is noted that following the developed post-melting procedure the encapsulation of the higher refractive index glass MW inside the host glass is realised while the MW remains intact, despite its lower $T_g$ when compared to that of the host glass. The reason behind this lies on the fact that the cylindrical MW is originally placed on the surface of the host glass block (Fig.1b). The latter glass is heated near to its $T_g$ to gain viscosity and facilitate the immersion of the MW. However, due to the shape and size of the MW, the contact area to the host glass is minimal and thus the heat transfer to the MW is inefficient to exceed the 148 °C that would make the MW melt. Moreover, the MW is ventilated from the top. This was the case for all fabricated samples upon visual inspection during the encapsulation procedure, while evidence of this appears also on the SEM image of Fig. 2f, that depicts the end face of the incorporated $0.3AgI+0.7AgPO_3$ glass MW remaining intact and exceeding the edge of the binary $AgPO_3$ host glass. However, in the interior of the host glass the softer glass MW would melt, creating a horizontal pathway of silver-rich higher refractive index glass within the lower refractive index host glass, as depicted schematically in Fig. S3.

The outcome of forming the silver-rich pathway along the $AgPO_3$ glass in terms of waveguiding becomes immediately apparent in Fig. 3. Fig. 3a shows an optical microscopy photo of the studied waveguide device, whereas Figs. 3b-h present the waveguide features of the green (Figs. 3b-d), blue (Figs. 3e-f), and red (Figs. 3g-h) light upon coupling in the beams of the corresponding cw lasers. The employed input powers are 1.2 mW, 1.2 mW, and 0.8 mW, for green, blue and red lasers, respectively. The first photo of each colour depicts the devices with lab lights on, whereas the other show the devices under dark conditions. Upon achieving a suitable incident light angle to couple light in the higher refractive index silver-rich channel the light propagates through the host glass, towards the other side of the device. Indicatively, inspection of Figs. 3d and f reveal bright spots on the end-face tip that was left to exceed the exit of the waveguide device (Fig. 2f).

In order to couple light in the waveguide efficiently, the focused incident beam was moved through an objective mounted on a x,y,z micrometric stage (Fig. S2), while the glass was slightly tilted in front of the objective, so that an incident light angle is formed. To demonstrate this, Fig. S4a presents an example of the red laser beam focused outside the waveguide pathway. The red and blue dotted lines in the same figure, point out that some of the light continues uncontrollably to the other side of the glass, i.e. without waveguiding features. On the contrary, once the beam is moved via the micrometric stage within the waveguide core, it waveguides through it, and deviates from the vertical orientation with respect to the objective (Fig. S4b). Similarly, Vid. S2 illustrates the coupling-in light process for the green laser. At the start of the video (first two seconds), the green laser beam is focused outside the waveguide pathway, and no light transmits through the glass to the other side of the waveguide. As the video progresses, the beam is moved towards the waveguide channel. Once the beam reaches the entrance point at three seconds onwards, the light propagation occurs, as it can be seen from the luminescence at the other side of the waveguide device. In both red and green light demonstrations, the glass is positioned exactly at the same place throughout the process, i.e. stacked on a microscope slide with double sided tape (Fig. S4).

The required incident light angle for efficient light transmission through the fabricated waveguides is rationalized in terms of the total internal reflection (TIR) principles [28,29]. In general, TIR is the phenomenon in which light that is travelling from a denser medium to a rarer medium is reflected in the former medium at the interface of the two media (Fig. S3). However, there are two necessary conditions for the occurrence of TIR between two glass optical media, as is the case in optical fiber architectures. First, the light must be travelling from a denser medium to a sparser medium, i.e. the refractive index of the former glass must be higher than that of the latter. Second, the light angle of incident must be greater than the critical angle that is determined from Snell's law [28,29]. Notably, both conditions are valid in the case of the studied waveguides. In particular, the refractive index of the denser silver-rich channel would be of 1.90 [27], whereas the corresponding value for the host glass that surrounds the waveguiding pathway is of 1.79 (Fig. S3) [27]. Moreover, the critical angle for

TIR in the so-formed waveguides is determined equal to 70.4°. As depicted in Fig. 3, the employed incident light angles are greater than the critical angle of 70.4°. Furthermore, to provide a reference in terms of the optical losses of the fabricated waveguide devices, we have studied extensively the output power of light with respect to the power of coupled in light upon employing the cw green laser. Fig. S5 presents the determined optical losses for various applied powers. The average transmission loss was determined equal to 6 db/cm, whereas the maximum obtained loss of 8 db/cm was also considerably low.

Along similar lines, Fig. 4 presents the operation of the multipath waveguides upon simultaneously transmitting green and red light. Figs. 4a-d depict a device in which two MWs are incorporated parallel to each other, whereas Figs. 4e-h show a waveguide glass in which the second MW is encapsulated in a diagonal direction, i.e. facing on the lower hand side of the host glass square rather than the opposite end face. Inspection of the optical microscope photos of the devices shown in Figs. 4a and 4e reveal the obvious marks of the two immersed MWs within the host glasses. The corresponding SEM photos of the multipath waveguide devices are shown in Figs. S6a and b. Fig. 4b presents the single red-light transmission within one MW, whereas as Figs. 4c and d depict the corresponding photos in which both red and green light are waveguided within the parallel pathways of the device, upon coupling light in from the opposite directions (Fig. S2). On the same manner, Fig. 4f depicts the single transmission of red light through the diagonally formed pathway, and Figs. 4g and h show the corresponding photos in which both green and red light are waveguided through the glass. Finally, Figs. S6c and d depict photos in which only the green laser is coupled in the devices. Overall, the captured photos demonstrate explicitly the two distinct waveguiding pathways formed following the encapsulation of the two MWs within the host glass.

The optical and structural properties of the employed glasses were also assessed. Fig. 5a presents the optical absorbance of the $AgPO_3$ host glass and the splat-quenched $0.3AgI+0.7AgPO_3$ glass that was used for drawing the MWs. For the sake of comparison, a corresponding spectrum of a fast-cooled $AgPO_3$ splat-quenched glass is also shown. Inspection of Fig. 5a reveals characteristic absorbance peaks at the 400 to 600 nm range for the host and the MW glasses, attributed to the presence of silver nanoparticles (AgNPs). Notably, the profile of the latter glass is more intense and shifted to longer wavelengths indicative for the formation of more and with wider size distribution AgNPs, upon the introduction of AgI [30]. On the contrary, such profile is absent from the corresponding profile of the ultrafast cooled glass, in which the formation of AgNPs is suppressed due to the rapid cooling.

The effect of the AgNPs presence on the structure of the employed phosphate glasses was investigated by Raman spectroscopy. Fig. 5b depicts the normalized Raman spectra of both $AgPO_3$ and $0.3AgI+0.7AgPO_3$ glasses in the forms of splat-quenched discs and drawn MWs. It is known that the metaphosphate network consists mainly of chains upon connecting phosphate tetrahedral units with bridging and non-bridging (terminal) oxygen atoms [24,26]. The obtained two main Raman features of Fig. 5b originate from these structural units. Indeed, the key band at 1140 cm$^{-1}$ is attributed to the symmetric stretching vibration of the terminal $PO_2^-$ groups, whereas the broader band at 675 cm$^{-1}$ arises from the symmetric stretching of the P-O-P bridges of the phosphate backbone [24,26]. The relative intensities of the two bands provides direct evidence on the population of the phosphate entities, and thus, on the phosphate network modifications. However, the analysis of Fig. 5b reveals no significant changes in the relative intensities of the Raman features between splat-quenched glasses and drawn MWs. Thus, it is concluded that despite the MW drawing process and the introduction of AgI the phosphate network is practically intact, rendering the employed glasses suitable for the fabrication of the studied waveguide devices.

STEM offers useful information on the morphology of the so-formed AgNPs within the host glass and the embedded MWs. Fig. 5c presents an indicative STEM photo of the AgNPs within the host glass, whereas Fig. 5d shows the corresponding image of AgNPs within the MWs glass. Inspection of the two samples reveals that the addition of AgI in the latter glass causes the agglomeration of AgNPs and the formation of larger size nanoclusters and particles. Data analysis of the AgNPs size distribution within the host $AgPO_3$ glass (Fig. 5c) indicates an average size of 20 nm. Rather differently, once the AgI is introduced for the MWs glass (Fig. 5d), the average AgNPs size increase up to 40 nm, while silver agglomeration results to for the formation of clusters up to 500 nm (Fig. 5e). Figs. 5f-j depict the elemental mapping of the sample by means of HRTEM, corresponding to the combined and to the individual maps for Ag, O, P, and I. The highlighted cluster regions in Fig. 5e, exhibit dominant Ag presence, whereas in the same regions significant absence of the other elements is noted. The presence of AgNPs within the silver-rich waveguide pathway of the fabricated devices, accounts plausibly for the obtained low losses (Fig. S5). It is well known that AgNPs absorb and scatter light with great efficiency. Indeed, they are often used as active centres for improving light absorption and transmission in various types of devices, that include photovoltaics [31], and fibers for random lasers and light emitting diodes [32]. In particular, AgNPs interact strongly with light since the conduction electrons on the metal surface exhibit a collective oscillation after light excitation, known as localized surface plasmon resonance (LSPR). The existence of

LSPR along with the matching of emission in the visible spectra range, amplifies the waveguided emission of light, and minimizes optical loss throughout the waveguide [32]. Apart from this, the presence of smaller and less AgNPs within the host glass block, accounts plausibly for the obtained random scattering points that can be seen in Figs. 3b and f, outside the silver-rich pathway of the waveguide.

## 4. Conclusions

We have shown the fabrication of advanced waveguides upon incorporating silver-rich glass MWs within a phosphate glass host, by means of a simple, fast, post-melting encapsulation procedure. The glass MWs were drawn from typical splat-quenched glasses, while exhibiting higher refractive index than the host glass. Upon embedding them within the host glass, a pathway of higher refractive index is formed, creating suitable conditions for total internal reflection across the waveguide. Moreover, the developed protocol allows the incorporation of multiple MWs within the same host, allowing simultaneously guiding features of different light sources and at different directions. Overall, we believe that the proposed method poses important technological advancements and opens the doors for the industrial scale production of glass waveguides and composites glasses for various photonic circuits, optoelectronic, and smart sign applications.

# Figures

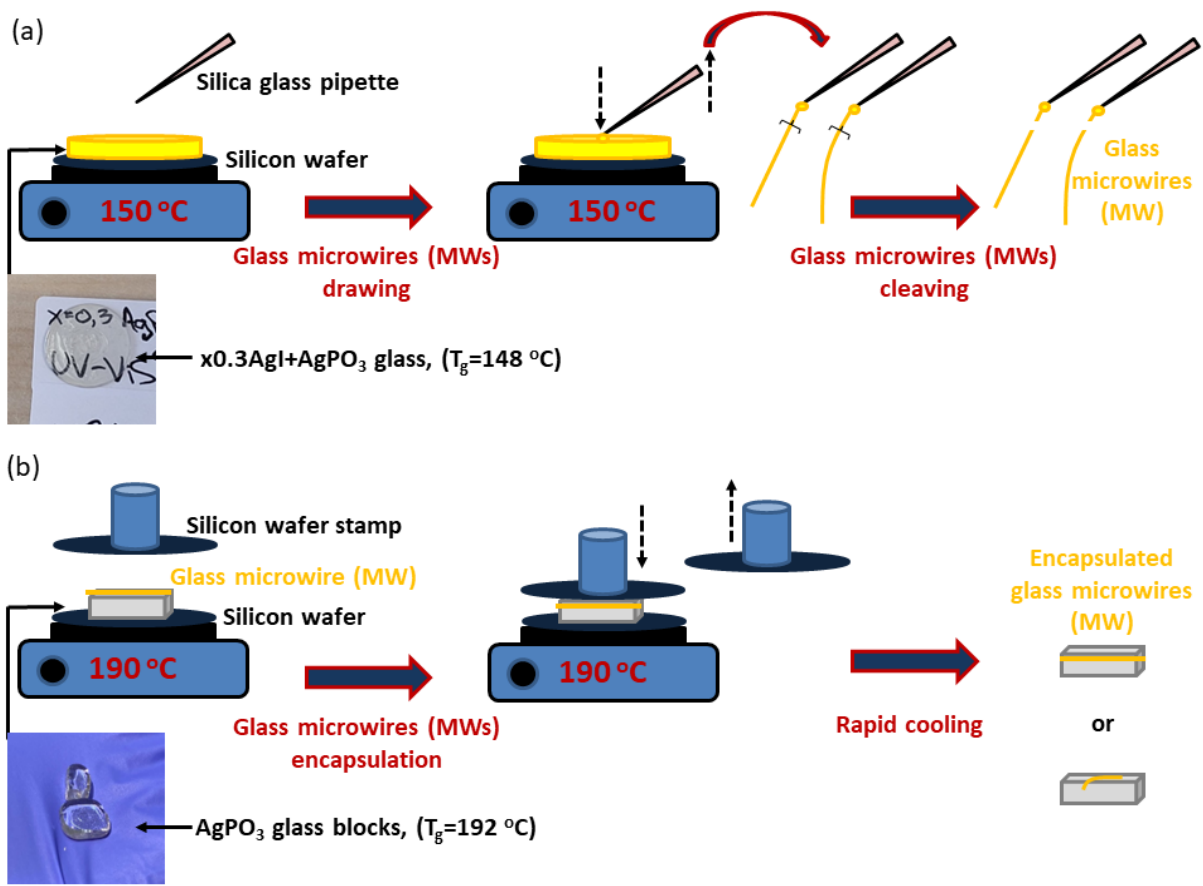

**Fig. 1: (a)** Schematic representation of the glass microwires (MW) drawing procedure from 0.3AgI+0.7AgPO3 splat-quenched glass that is depicted in the inset photograph. **(b)** Schematic representation of the MWs encapsulation within the host AgPO3 glass. Indicative host glass prisms are depicted in the inset photograph.

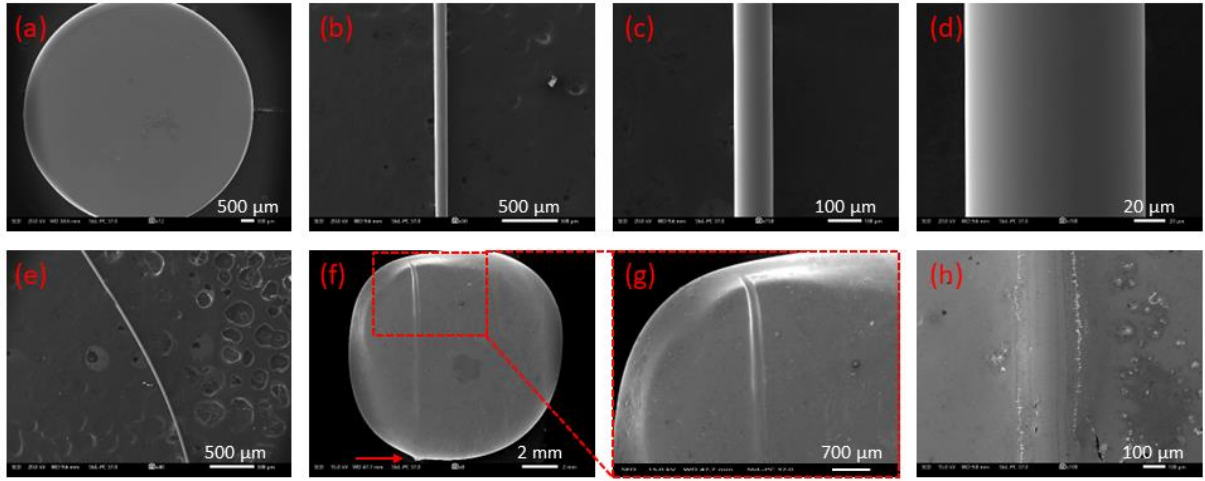

**Fig. 2: (a)** Scanning electron microscopy (SEM) images of the 0.3AgI+0.7AgPO$_3$ splat-quenched glass. **(b)** Typical SEM image of a glass MW drawn from the splat-quenched glass. **(c) and (d)** Higher magnification images of the same glass MW. **(e)** SEM image of a curved MW. **(f)** SEM image of a single MW waveguide device upon encapsulation of the MW within the host glass. **(g)** Magnified area of the device. **(h)** Magnified area of the MW immersion point on the surface of the host glass.

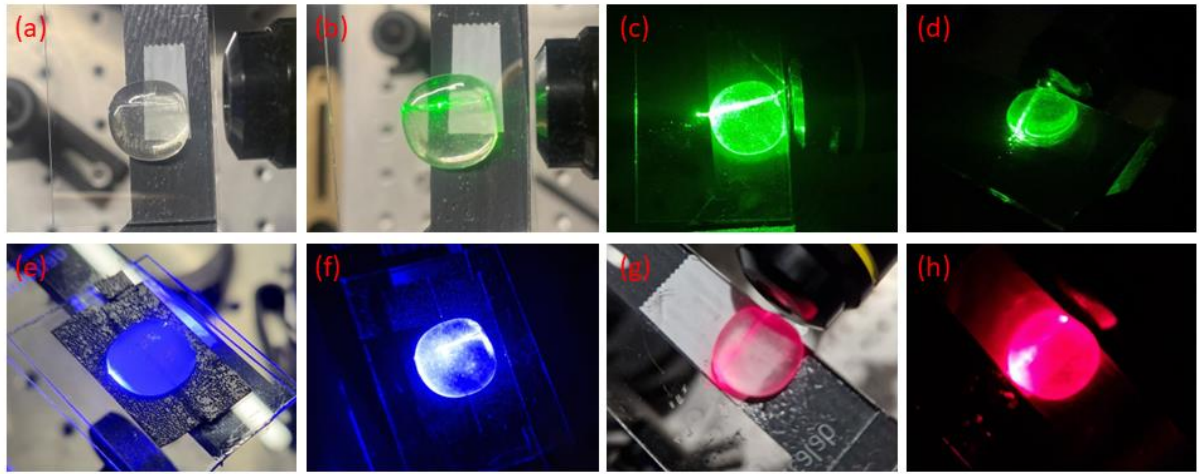

**Fig. 3: (a)** Optical microscope photo of a typical single MW waveguide device, in which the MW immersion point is visible. Waveguiding features of the green laser with lights-on **(b)**, and under dark **(c)**. **(d)** Picture of the same waveguide device under different angle, so that the output point is more visible. Waveguiding features of the blue laser with lights-on **(e)**, and under dark **(f)**. Waveguiding features of the red laser with lights-on **(g)**, and under dark **(h)**.

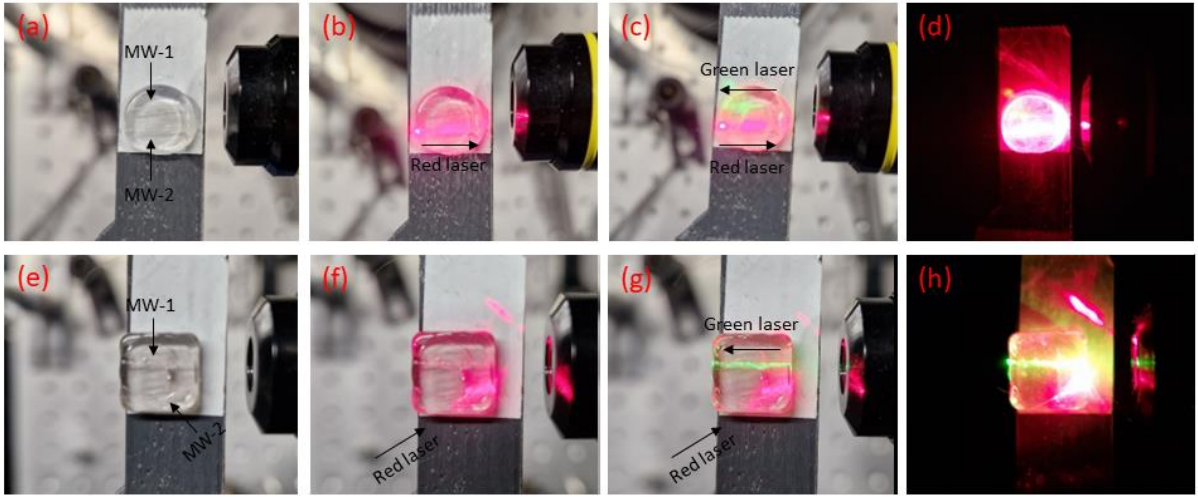

**Fig. 4: (a)** Optical microscope photo of a multipath waveguide, in which two parallel MWs are incorporated. The arrows point out the two immersion points. **(b)** Red laser waveguiding throughout one of the incorporated MWs (left to right direction). **(c)** Red and green (right to left) laser waveguiding throughout the two parallel MWs. **(d)** Same as previous under dark. **(e)** Optical microscope photo of a multipath waveguide, in which the second MW is encapsulated diagonally to the other. **(f)** Red laser transmission through the diagonally placed MW (bottom to right direction). **(g)** Red and green (right to left) light transmission through the diagonally and parallel placed MWs. **(h)** Same as previous under dark.

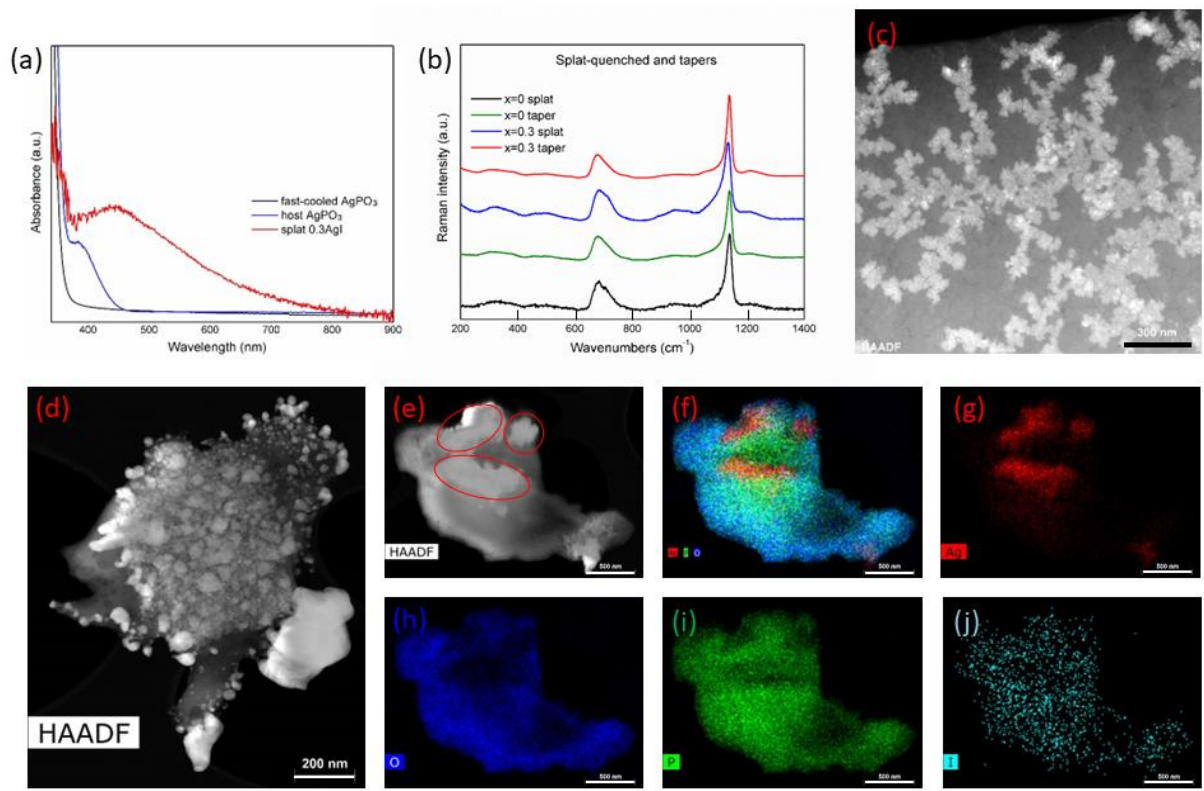

**Fig. 5:** Optical absorbance **(a)**, and Raman **(b)** of the employed glasses. Scanning transmission electron microscopy (STEM) photo of the silver nanoparticles (AgNPs) within the host $AgPO_3$ glass **(c)**, and the $0.3AgI+0.7AgPO_3$ MWs glass **(d)**. **(e)** HAADF image of the agglomerated AgNPs within the MWs glass. **(f)** HRTEM-EDX elemental mapping of the same sample, along with the individual element spatial analysis for Ag **(g)**, O **(h)**, P **(i)**, and I **(j)**.